# Evolution of the surface atomic structure of multielement oxide films: curse or blessing?

Giada Franceschi,*[a] Renè Heller,[b] Michael Schmid,[a] Ulrike Diebold,[a] and Michele Riva[a]

**Atomically resolved scanning tunneling microscopy (STM) and x-ray photoelectron spectroscopy (XPS) are used to gain atomic-scale insights into the heteroepitaxy of lanthanum-strontium manganite (LSMO, $La_{1-x}Sr_xMnO_{3-\delta}$, $x \approx 0.2$) on $SrTiO_3(110)$. LSMO is a perovskite oxide characterized by several composition-dependent surface reconstructions. The flexibility of the surface allows it to incorporate nonstoichiometries during growth, which result in composition-dependent surface atomic structures. This happens up to a critical point, where phase separation occurs, clusters rich in the excess cations form at the surface, and films show a rough morphology. To limit the nonstoichiometry introduced by non-optimal growth conditions, it proves useful to monitor the changes in surface atomic structures as a function of the PLD parameters and tune the latter accordingly.**

## 1. Introduction

Perovskite oxides dominate a host of established and emerging technologies due to their extraordinary tunability.[1–4] Characterized by the chemical formula $ABO_3$, perovskites and related crystal structures can accommodate about 30 elements on the A site and over half the periodic table on the B site.[5] This opens up the attractive possibility to control the interplay between spin, charge, orbital, and lattice degrees of freedom and achieve unique properties.[3] One example is $La_{1-x}Sr_xMnO_{3-\delta}$ (lanthanum–strontium manganite, LSMO), which shows doping-dependent transitions from metal to insulator and from (anti)ferro- to paramagnetic, as well as interesting catalytic properties.[6–10]

Because of their high sensitivity to stoichiometry and crystal structure changes,[11] the properties of perovskite oxides are best explored by working with single-crystalline samples or bulk-like epitaxial thin films. The growth of ideal perovskite-oxide films is challenging, however. In pulsed laser deposition (PLD) – the technique of choice for multielement oxides – many parameters can affect preferentially one element or another in the compound. These parameters include the laser energy density, spot size, laser focal spot on the target surface, pressure and nature of the ambient gas, and deposition rate.[12–15] Depending on the precise values of the deposition parameters (sometimes hard to reproduce in different laboratories[13,16,17]), elements may be preferentially ablated at the target, preferentially scattered by the background gas, and have preferential sticking to the substrate.[13] Put simply, stoichiometric targets do not warrant stoichiometric films. The crystallinity, morphology, and other macroscopic properties are influenced together with the composition, and the effects are more pronounced at thicknesses larger than a few nanometres. LSMO films exemplify the struggle. Their morphology, composition, transport, and magnetic properties are highly sensitive to the growth conditions.[16,18–24] Crystalline precipitates during nonstoichiometric growth are common.[25,26]

The causes for morphological roughening are numerous and intertwined. In the simple case of one-component films, deposition rate, energetics, attachment kinetics at step edges, mechanical stress, angle dependent rate, capillarity, viscous flow, and nucleation are known to be relevant.[27–32] These effects are expected to play a role also within the growth of perovskite oxides. This work demonstrates that additional effects – not considered in traditional models and related to the surface atomic details of the growing films – are important for understanding and controlling the complex growth behaviours of perovskite oxides.

Perovskite oxides are known to exhibit a host of composition-dependent atomic-scale surface structures (also named surface reconstructions).[33–40] Previous studies on $SrTiO_3(110)$ have already shown that these reconstructions are critically important during epitaxial film growth. If the deposition parameters are not optimized and the growth is nonstoichiometric, excess cations can segregate to the film surface[39] and alter its atomic structure. The nonstoichiometry of the deposited material might be minute (in the order of 0.05 layers, or about $10^{13}$ cm$^{-2}$) but its influence on the surface atomic structure can be detected by reflection high-energy electron diffraction (RHEED),[41] low-energy electron diffraction (LEED), and scanning tunneling microscopy (STM).[39] Importantly, these changes can alter the growth mechanisms[42,43] and the surface morphology: If reconstructions with different sticking properties develop and coexist on the surface, pits might develop on the low-sticking areas.[13]

The existence of surface reconstructions does not solely bring negative implications. As it was shown for $SrTiO_3$,[39] it also opens up attractive opportunities. In cases where all the introduced non-stoichiometry segregates to the surface of the growing film, optimal growth conditions and stoichiometric growth with virtually unlimited accuracy can be achieved by monitoring the changes in the surface

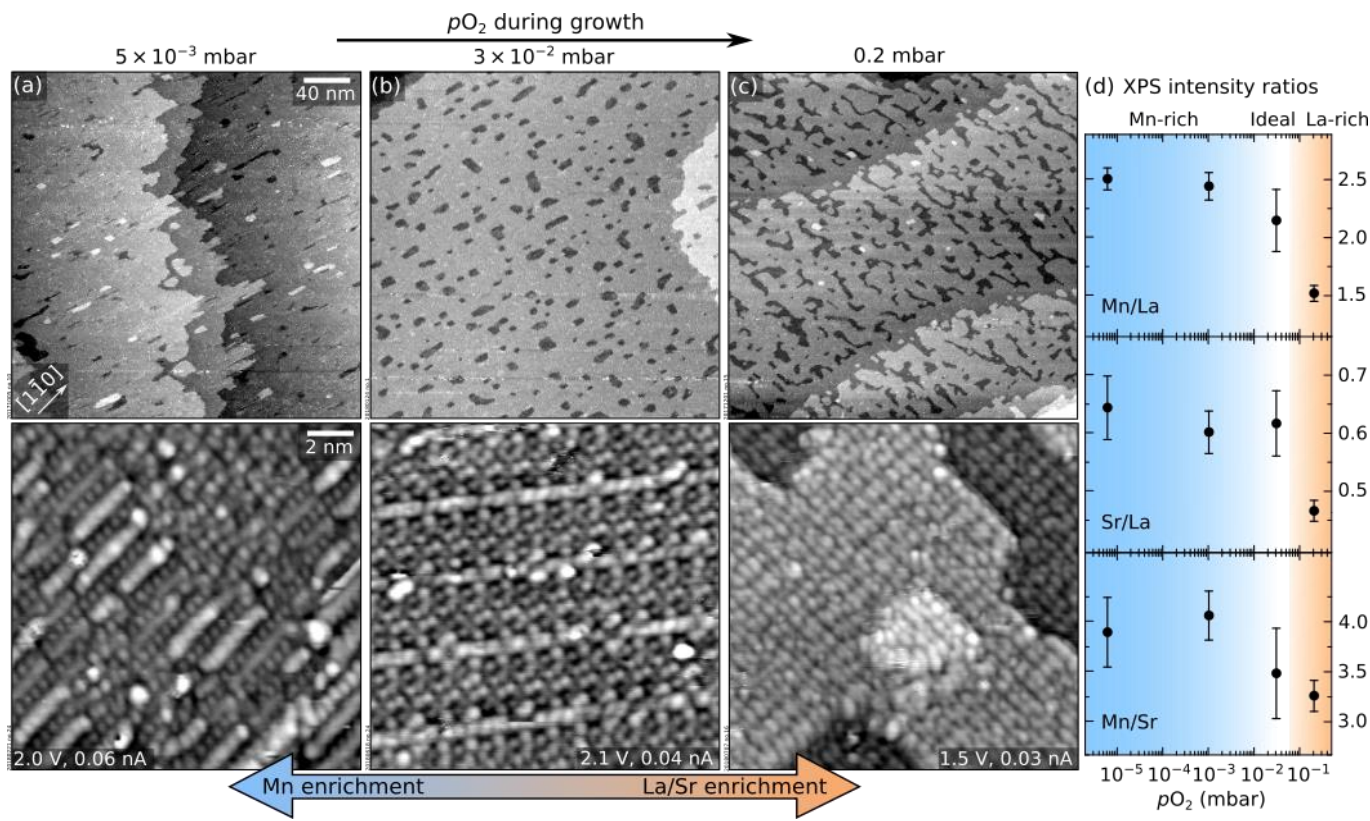

**Fig. 1.** Effect of $p_{O_2}$ during PLD on the morphology and composition of LSMO(110) films. (a–c) Top row: 300 × 300 nm$^2$ STM images ($V_{sample}$ = +2 V, $I_{tunnel}$ = 0.2 nA) of thin films (9 layers, ≈ 2.5 nm thickness) grown at 1 Hz, 1.9 J/cm$^2$, 700 °C, and 5 × 10$^{-3}$ mbar ≤ $p_{O_2}$ ≤ 0.2 mbar; bottom row: 15 × 15 nm$^2$ STM images, high-pass-filtered for displaying purposes. (d) XPS intensity ratios of Mn 2$p$, La 4$d$, and Sr 3$d$ peaks as a function of $p_{O_2}$; lower $p_{O_2}$ yields Mn-richer surfaces (the analysis includes an additional datapoint of a 9-layer-thick film grown at 5 × 10$^{-6}$ mbar). Note that the Sr signal originating from the substrate cannot be decoupled from the one in the film.

atomic structure as a function of film thickness and deposition parameters.[39]

This work builds on previous studies on SrTiO$_3$ homoepitaxy. It focuses on LSMO films deposited on well-defined SrTiO$_3$(110) substrates (see Section S1 of the ESI† for details on the setups and the growth). It investigates how the surface atomic structures of LSMO evolve during growth as a function of different parameters and film thicknesses, and how to leverage such changes to optimize the deposition parameters. The concepts are showcased by depositing under controlled conditions, where one parameter at a time is carefully modified – in this case, the value of the O$_2$ background pressure ($p_{O_2}$) within the incongruent transfer regime, where the lighter Mn species scatter more than the heavier La and Sr, leading to enrichment in La and Sr (Mn) at higher (lower) O$_2$ pressures. The surface evolution is monitored with XPS and STM. Akin to SrTiO$_3$, non-stoichiometries shift the surface atomic structure along established surface phase diagrams, following their substantial segregation to the surface.[40,44] When the excess material cannot be accommodated by a suitable surface reconstruction, it precipitates in the form of clusters rich in the excess cations. Nonetheless, appropriate ultrahigh vacuum (UHV)-based surface treatments can heal the surface. A method is presented to optimize the PLD parameters to grow high-quality multielement films. Different from typical practices, it does not rely on *ex-situ*, *post-mortem*, bulk analyses. Rather, it leverages the changes in surface atomic structure following the partial segregation of non-stoichiometry to adjust the PLD parameters. *Ex-situ* bulk analyses are offered to support the quality of the films and provide a link with standard characterization techniques.

## 2. Insights into the growth of LSMO(110)

This section addresses the role of $p_{O_2}$ during PLD on the morphology and composition of LSMO(110) films. Figures 1a–c show the STM morphology of three thin films grown at $p_{O_2}$ values ranging between 5 × 10$^{-3}$ mbar and 0.2 mbar (all other parameters are nominally the same: 1 Hz laser repetition rate, 1.9 J/cm$^2$ laser fluence, 700 °C substrate temperature). All films have the same thickness of ≈ 2.5 nm or 9 layers, where one layer corresponds to the separation between two (110) planes, ≈ 0.276 nm.

In Fig. 1, all films appear atomically flat on the scale of a few hundred nanometres (top row). On a smaller scale (bottom row), different surface structures with different periodicities (i.e., different surface reconstructions) are evident. The same reconstructions were observed by depositing sub-monolayer amounts of Mn on a stoichiometric film in order to establish a quantitative surface phase diagram of LSMO(110).[44] Comparing this phase diagram with Fig. 1 reveals that the films grown at lower pressures exhibit reconstructions richer in Mn. This is confirmed by the XPS analysis of

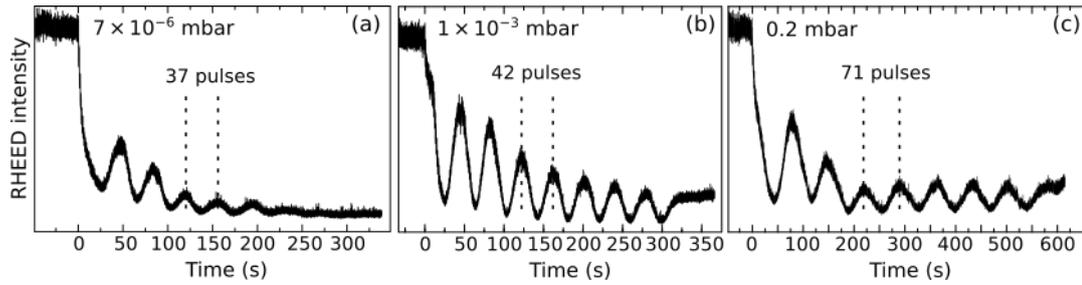

**Fig. 2.** RHEED oscillations during the growth of thin LSMO(110) films (≈ 2.5 nm-thick) at different $p_{O_2}$ (values indicated in the respective panels) and otherwise identical growth parameters (1 Hz, 1.9 J/cm$^2$, 700 °C). Note that the data do not correspond to the same films as shown in Fig. 1.

Fig. 1d. Note that the variation of the Mn/La XPS intensity ratio (Fig. 1d) by more than a factor of 1.6 is much higher than what would be possible in perovskite-type LSMO according to the bulk phase diagram.[45] While STM shows different surface reconstructions (Fig. 1a–c), all these surface phases are based on the same perovskite lattice of the underlying layers.[44] There is no evidence for the formation of different phases in the deeper layers of these films. This indicates that the stoichiometry variations at the surface are larger than in the deeper layers of the film.

Figure 2 displays the intensity of the specular RHEED spot versus the deposition time for three $p_{O_2}$ values. All depositions occur in a layer-by-layer mode (one RHEED oscillation equals one layer). Higher pressures produce longer periods, an indication that less material reaches the substrate. Moreover, the intensities of the minima and maxima differ for the three films. This is not necessarily always an indication for better or worse layer-by-layer growth, however. The different surface reconstructions developing during growth (Fig. 1) could potentially cause diffraction conditions different from the starting point, which may explain some amplitude variations of the intensity oscillations.[13] Nevertheless, the more pronounced decay observed in Fig. 2a probably indicates the slightly rougher surface morphology of the corresponding film (observed in STM, not shown here).

Growing thicker films at the same conditions as in Figs. 1a–c induces dramatic morphology changes, see Fig. 3. The lowest and highest pressures (5 × 10$^{-3}$ mbar, Fig. 3a; 0.2 mbar, Fig. 3c) produce new features a few nanometres in height, located mainly at the step edges and poorly conductive, as judged by the behaviour of STM. At the intermediate pressure of 4 × 10$^{-2}$ mbar, the surface preserves its flatness up to a thickness of 132 nm. Note that the imperfect films shown in Fig. 3 are significantly thinner: Clusters appears at thicknesses of only 5.5 and 11 nm in Figs. 3a and 3c, respectively.

The bulk properties of films without clusters were characterized with RBS and XRD, see Figs. S1 and S2, ESI.† The films are stoichiometric and crystalline. RBS quantified the composition as (La$_{0.78 \pm 0.03}$Sr$_{0.22 \pm 0.03}$)$_{1.06 \pm 0.05}$MnO$_3$, close to the target's (La$_{0.79 \pm 0.02}$Sr$_{0.21 \pm 0.02}$)$_{0.96 \pm 0.08}$MnO$_3$. The analysis of XRD reciprocal-space maps reveals that the films are crystalline but only partially relaxed – by 34.7% along [$1\bar{1}0$] and 6% along [001]. This is expected for heteroepitaxial films under slight stress[46] that relax by introducing misfit dislocations and forming mosaics. The residual deformation in the film in all three directions has an absolute value of less than 2 pm. Best-fit lattice constants and angles are reported in Table S1 of the ESI.†

## Discussion

As mentioned in the introduction, stoichiometric growth in PLD can be achieved following the optimization of many parameters,[12] among others the oxygen background pressure and the laser fluence.[47] As discussed below, both parameters are responsible for the growth behaviours summarized by Fig. 1 and Fig. 3.

Changes in the film composition as a function of $p_{O_2}$ in multi-element oxides can be explained within the three-pressure-regimes framework[23,48,49] (regardless of the plume composition right after the ablation, which can be affected by the laser fluence, see below). At low-enough $p_{O_2}$, the ablated species are congruently transferred to the substrate. At intermediate pressures, lighter species are scattered more than heavier ones and the film becomes enriched with the heavier species as $p_{O_2}$ increases.[50] At very high $p_{O_2}$, in the so-called shock-wave regime, all ablated species are slowed down equally; they are kept confined in the plume and transferred congruently to the substrate. These pressure regimes are identified not only by $p_{O_2}$ but also by the target-to-substrate distance, $D$,[16] (for the experimental setup used here, $D$ = 55 mm). Here, the low-pressure regime occurs at $p_{O_2} \leq 5 \times 10^{-3}$ mbar: Below this value, there is no change in the XPS signals in Fig. 1, indicating congruent transfer. The intermediate pressure regime occurs at $3 \times 10^{-2} \leq p_{O_2} \leq 0.2$ mbar: here, the Mn content decreases with increasing pressure (Mn is the lightest cation in LSMO).

If the laser fluence is chosen such that all the species are ablated congruently at the target, the target has ideal stoichiometry, and the sticking probability of all ablated species on the surface is the same, it is possible to achieve near-ideal stoichiometries by depositing within either the low- or the high-pressure regimes. However, it is rare that all these conditions are fulfilled. Tiny deviations in the laser fluence or pulse duration can affect the ablation significantly, inducing the preferential ablation of one element over another in the multielement oxide. This is exemplified by the homoepitaxy of SrTiO$_3$, where congruent ablation is achieved only within a very narrow window of laser fluence.[13,17,39,51] Moreover, reproducing the laser fluence in different PLD setups is challenging:[13] The spot size and the intensity distribution within the spot affect the deposition greatly;[12] the most common way to adjust the UV pulse energy, by changing the discharge voltage of the UV laser, affects the pulse duration and beam divergences. Moreover, UV laser gases age over time, causing increasing pulse-to-pulse standard deviations and ill-defined fluences.

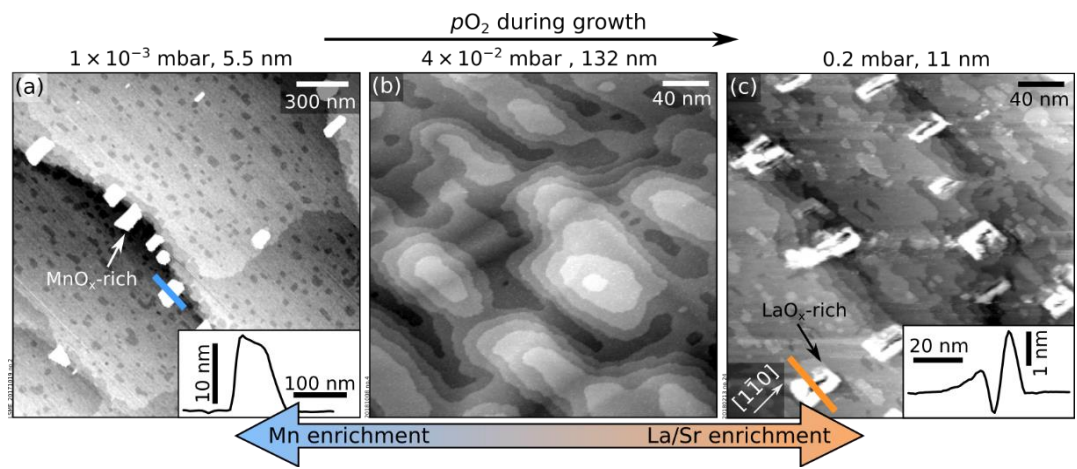

**Fig. 3.** Accumulation of nonstoichiometry at the surface of LSMO(110) films. (b, c) 300 × 300 nm$^2$ STM images and (a) 2.0 × 2.0 μm$^2$ AFM image of LSMO(110) films thicker than those in Fig. 1 grown at different $p_{O_2}$ and otherwise identical deposition parameters as to Fig. 1. Nonstoichiometric growth (a, c) conditions result in poorly conductive, few nanometers-high features on the surface that are identified as manganese- and lanthanum-oxide-rich clusters at low and high $p_{O_2}$, respectively. Line scans over selected clusters are shown in the insets. (b) Ideal stoichiometry leads to films without precipitates even at a large thickness.

This work demonstrates how to achieve near-ideal stoichiometries even when the ablation is incongruent by exploiting the wider tunability window offered by the oxygen background pressure: The cation non-stoichiometry caused by preferential ablation can be mitigated by exploiting preferential scattering effects within the intermediate-pressure regime.

The "intermediate-pressure" film of Fig. 3b has a composition close to the that of the target (see RBS analysis in Fig. S1c, ESI†). Then, according to the XPS data in Fig. 1d, films grown at lower and higher $p_{O_2}$ (i.e., congruent transfer regimes) must be Mn rich and Mn deficient, respectively. This means that Mn is preferentially ablated at the target with the chosen laser fluence: Mn-rich films are obtained at low $p_{O_2}$ because the Mn-enriched plume is transferred congruently. The ideal stoichiometry is achieved at a specific value of intermediate $p_{O_2}$ where the excess Mn is scattered more than the heavy La species. Finally, at even higher pressures, more than the excess Mn is scattered away and the films grow La-rich. The preferential ablation of Mn at the target is possibly caused by the laser fluence being too low. Previous studies on SrTiO$_3$ homoepitaxy have shown that low fluences produce Sr-rich films,[52,53] probably because of the higher vapor pressure of Sr compared to Ti; similarly, Mn could be preferentially ablated at low laser fluences because of its higher vapor pressure compared to La.

The different film compositions translate into different atomic-scale surface structures (Fig. 1a–c). The changes in the RHEED patterns of LSMO films grown at different $p_{O_2}$ observed in the literature[22] likely arise from the different surface reconstructions formed in each regime.

During SrTiO$_3$(110) homoepitaxy, all non-stoichiometry segregates to the surface and changes its composition and atomic structure according to its composition phase diagrams.[13,39] While it is not possible to prove that such full segregation occurs on LSMO(110) as well, several pieces of evidence support that segregation occurs at least partially, as seen by: (i) different surface structures forming on thin films with different compositions (Fig. 1), (ii) the evolution of the surface structures with increasing thickness (see Section 3), (iii) the XPS intensities mentioned in the previous paragraph, and (iv) the formation of non-conductive clusters when a critical thickness is overcome (Fig. 3). Since LSMO is an electrical conductor at room temperature, these clusters must consist of a different material. Hence, they are reasonably assigned to MnO$_x$ and LaO$_x$ excess introduced within the low- and high- pressure regimes, respectively. This is supported by the fact that similar clusters are observed after depositing large amounts (above ≈2 ML) of Mn and La on well-defined LSMO(110) surfaces followed by short annealing times (≤ 20 min); and by the previously reported formation of MnO$_x$ precipitates in epitaxial LSMO(001) films under Mn-rich conditions.[25]

The 3D clusters form when the surface cannot accommodate the excess cations by modifying its atomic structure. It was already reported that Mn (La) species stick less on Mn(La)-richer surfaces.[44] A careful inspection of the LaO$_x$ clusters in Fig. 3c reveals that they actually consist of pits surrounded by tall rims – similar to the features formed during the Ti-rich homoepitaxy of SrTiO$_3$(110).[13] There, the formation of pits was assigned to surface-dependent sticking and diffusion effects. The case of LSMO(110) appears similar. At La-rich conditions, the surface structure shifts towards La-rich reconstructions;[44] at a critical composition, it becomes more favourable to nucleate and grow La-rich clusters than to incorporate more La in the surface structure.

Preferential sticking effects could act as a self-adjusting feedback mechanism for the film stoichiometry under slightly nonstoichiometric conditions: For slightly La-rich fluxes, the surface will gradually shift towards A-site richer reconstructions, onto which Mn sticks better, such that the surface shifts back towards Mn-richer structures. Now more La sticks, and so on. In principle, the forgiveness of this growth mechanism can allow growing atomically flat and stoichiometric films even under slightly nonstoichiometric conditions. However, as discussed above, this will work only up to a certain point: if the cation non-stoichiometry introduced by the incoming flux exceeds the capability of the surface to accommodate the non-stoichiometry via a change of the surface structure, oxide clusters form instead. One should also mention that, compared to $SrTiO_3(110)$, the surface reconstructions of LSMO(110) are separated by larger compositional differences.[39,44] Hence, larger deposited non-stoichiometry can be accommodated at the surface of LSMO(110) films while yielding atomically flat films over a larger window of growth parameters.

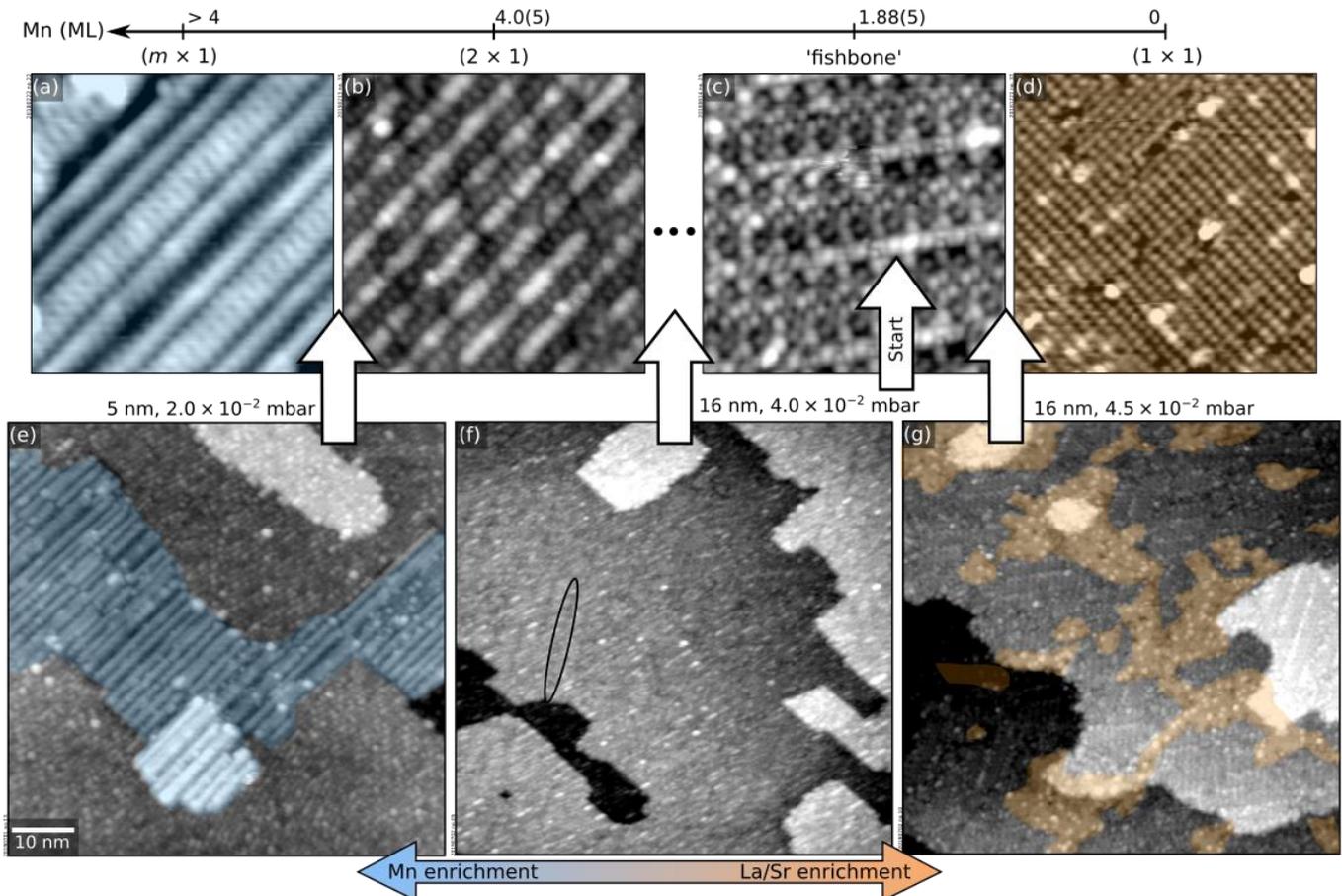

**Fig. 4.** Pinpointing the exact $p_{O_2}$ value for optimal LSMO(110) growth. (a–d) 12 × 12 nm² STM images of selected surface structures of the surface phase diagram of LSMO(110),[44] obtained by depositing controlled amounts of La or Mn from $La_2O_3$ and MnO targets in PLD plus annealing at 700 °C and 0.2 mbar $O_2$ (see Methods section in the ESI;† 1 ML corresponds to the number of Mn sites in an $(AMnO)_2$ plane of LSMO(110), i.e., 4.64 × 10¹⁴ at./cm²). (e–g) 70 × 70 nm² STM images of LSMO films of various thicknesses grown at different $p_{O_2}$, always starting from LSMO(110) films with the fishbone structure of panel (c). (e) 5-nm-thick film grown at 2.0 × 10⁻² mbar $O_2$, displaying patches of the Mn-rich (m × 1) structure of panel (a). (f) 16-nm-thick film grown at 4.0 × 10⁻² mbar $O_2$, with a surface reconstruction in between the fishbone and (2 × 1). (g) 16-nm-thick film grown at 4.5 × 10⁻² mbar $O_2$, with patches of the (1 × 1) and of the fishbone reconstructions.

One expects non-stoichiometry to accumulate at surfaces rather than in the bulk when forming bulk defects is comparatively more costly.[39,54] Interestingly, the bulk phase diagrams of $La_{0.8}Sr_{0.2}MnO_3$[45] show that cation excesses are not easily incorporated in the bulk at UHV-compatible pressures. At atmospheric pressure, the system can accommodate significant excess of oxygen (and, possibly, cations). Nonetheless, our data show that excess Mn tends to float to the surface and change the surface atomic structure. Thus, non-stoichiometry segregation appears to be a powerful effect even in those perovskite oxides whose phase diagram would allow creating bulk defects.

## 3. Strategies to obtain ideal films

This section illustrates a strategy to pinpoint the optimal growth conditions for LSMO(110) films (in this case, the value of $p_{O_2}$ required to create a stoichiometric film in spite of the Mn-rich ablation). Several films were grown within the optimal range around 10⁻² mbar (see Section 2). The corresponding surface structure changes were monitored with STM. Figure 4 summarizes the results. For reference, the top row of Fig. 4 reports the surface phase diagram of LSMO(110) in the relevant $p_{O_2}$ range.[44] Before each deposition, the surface was prepared to exhibit the 'fishbone' reconstruction of Fig. 4c, i.e., a $\begin{pmatrix} \pm 4 & 5 \\ \mp 1 & 2 \end{pmatrix}$ superstructure described in detail in Ref. [44], which introduces the shorthand notation for this and other reconstructions.

First, 5 nm were grown at 2.0 × 10⁻² mbar $O_2$. The surface (Fig. 4e) is atomically flat but exhibits patches of the (m × 1) structure of Fig. 4a, indicating that this pressure introduces a significant Mn

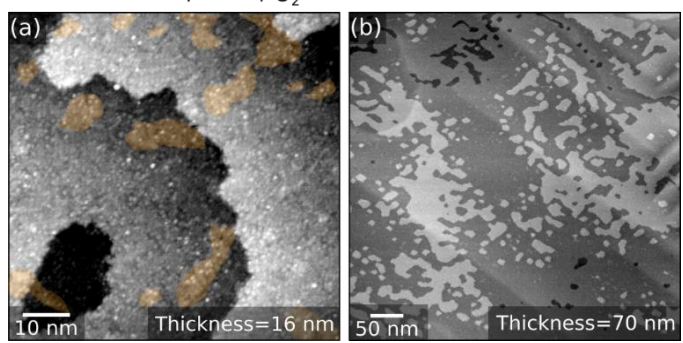

**Fig. 5.** LSMO(110) film grown at optimized conditions. (a) 70 × 70 nm$^2$ and (b) 500 × 500 nm$^2$ STM images of 16-nm-thick and of 70-nm-thick LSMO(110) films, respectively, grown onto a fishbone-reconstructed LSMO(110) surface at 4.2 × 10$^{-2}$ mbar O$_2$. Both morphologies are atomically flat. At the atomic scale, (1 × 1) patches are visible [orange in panel (a)].

excess. Indeed, MnO$_x$-rich clusters are formed at 10 nm thickness (not shown).

As learned from Fig. 1, higher values of $p_{O_2}$ should introduce less Mn. By growing a film of 5 nm thickness at 4.0 × 10$^{-2}$ mbar, the surface structure remains essentially the same (not shown). Nonetheless, increasing the thickness to 16 nm (Fig. 4f) reveals a mix of the fishbone reconstruction (solid oval) and a reconstruction between the fishbone and the (2 × 1),[44] indicating that the conditions are still Mn-enriching. Thus, the pressure should be increased further.

Growing 16 nm at $p_{O_2}$ = 4.5 × 10$^{-2}$ mbar on SrTiO$_3$ pushes the surface toward the other end of the surface phase diagram (Fig. 4g): Patches of the (1 × 1) reconstruction of Fig. 4d (A-site richer than the fishbone) coexist with fishbone-reconstructed areas. However, these conditions are still not ideal: Larger thicknesses produce increasing areal coverages of (1 × 1) followed by the formation of AO$_x$-rich clusters.

The ideal condition is reached at 4.2 × 10$^{-2}$ mbar. A film of 16 nm thickness shows a small (1 × 1) coverage (Fig. 5a). A flat surface is maintained up to 70 nm thickness (Fig. 5b).

This section has shown how to grow flat films by detecting the non-stoichiometry from its influence on the surface reconstruction and adjusting $p_{O_2}$ accordingly. Notably, it is also possible to recover films with precipitates by means of appropriate UHV treatments. Section S3 of ESI† shows that alternating annealing at oxidizing and reducing conditions favors surface diffusion and the flattening of the surface (as shown in Ref. [55], this is a common behavior of oxide materials). This strategy is more effective than standard sputtering–annealing cycles.

## Conclusions

This work addresses the correlation between non-stoichiometry (systematically tuned by varying the O$_2$ background pressure), surface morphology, and surface atomic structures in PLD-grown LSMO(110) films by combining chemical analysis by XPS and atomically resolved STM. The often-overlooked surface atomic details have important implications for the growth. The surface can incorporate excess cations and affect the film composition through preferential sticking. The composition-dependent surface atomic structures also offer a precise metric to optimize the PLD parameters and achieve high-quality films. Since non-stoichiometries tend to segregate to the film surface and are prone to change its surface atomic structure, a stable atomic-scale surface structure at increasing film thicknesses indicates that the chosen PLD parameters yield a close-to-stoichiometric growth. On the other hand, a shift in the surface atomic structure indicates that cation excesses are introduced. To achieve close-to-stoichiometric films and avoid precipitation of undesired phases, the PLD parameters should be tuned such that the atomic-scale structure of the surface remains stable.

Monitoring the details of the surface atomic structures of the film also sheds light on previously disregarded mechanisms inducing morphological roughening. When the introduced non-stoichiometry exceeds a critical value, the surface cannot accommodate the excess cations by changing its atomic structure anymore. Instead, clusters of the excess material develop at the surface. In such cases, alternating annealing treatments at oxidizing and reducing conditions is an effective means to remediate the surface morphology.

Many phenomena observed during the growth of LSMO(110), including non-stoichiometry segregation that alters the surface structure, surface-dependent incorporation of deposited cations, and phase separation, seem to be a general trait of perovskite oxides and possibly many other multi-element compounds.[13,14,56,57] A growing number of systems is reported to exhibit composition-related surface reconstructions with an apparent tendency to accommodate cation excesses. The authors contend that the findings reported here reflect general behaviours of complex oxide films, independent of the growth technique. The insights and methods presented can guide the growth optimization of perovskite-oxide films.

## Author Contributions

Conceptualization: GF, MR. Investigation: GF, MR, RH. Validation: MR, MS, UD. Supervision: MR, MS, UD. Funding acquisition: UD. Writing, original draft: GF. Writing, review and editing: all authors.

## Conflicts of interest

There are no conflicts to declare.

## Acknowledgements

The authors were supported by the Austrian Science Fund (FWF) through project SFB-F81 "Taming Complexity in materials modeling" (TACO) and by the European Research Council (ERC) under the European Union's Horizon 2020 research and innovation programme (grant agreement No. 883395, Advanced Research Grant 'WatFun').

## Notes and references*

## Supplementary Information for:

## Evolution of the surface atomic structure of multielement oxide films: curse or blessing?

Giada Franceschi, Renè Heller, Michael Schmid, Ulrike Diebold, and Michele Riva

### S1. Materials and methods

The LSMO/SrTiO$_3$(110) films were grown in a ultra-high vacuum pulsed-laser deposition (UHV PLD) setup (base pressure < 4 × 10$^{-10}$ mbar after bake-out) fitted for high-pressure and high-temperature growth experiments. This PLD setup is attached via an intermediate UHV chamber to a surface characterization facility (base pressure below 4 × 10$^{-11}$ mbar) comprising scanning tunneling microscopy (STM), low-energy electron diffraction (LEED), x-ray photoelectron spectroscopy (XPS), and low-energy He$^+$-ion scattering (LEIS). The UHV system and all technical details about the growth of LSMO/SrTiO$_3$(110) films are discussed elsewhere.[1,2]

The SrTiO$_3$(110) substrates (single crystals from CrysTec GmbH, 0.5 wt.% Nb-doped, 5 × 5 × 0.5 mm$^3$, one-side polished, miscut < 0.3°) were prepared and characterized in UHV to exhibit a mixture of (4 × 1) and (5 × 1) surface reconstructions[3] (checked in STM; purity checked in XPS before growth).

The LSMO films were grown at 1 Hz laser repetition rate, laser fluence of 1.9 J/cm$^2$, 700 °C substrate temperature, and O$_2$ pressures between 7 × 10$^{-6}$ mbar and 0.2 mbar. The surface composition and atomic structure of some films used as a substrate for further deposition were tuned by PLD of controlled amounts of La and Mn from La$_2$O$_3$ and MnO targets (2 Hz, 1.5 J/cm$^2$, 0.2 mbar O$_2$, RT), followed by annealing for at least 30 min at 700 °C, 0.2 mbar O$_2$.[1] The thickness was estimated by monitoring the evolution of the specular-spot intensity of RHEED (in the layer-by-layer growth mode one oscillation corresponds to one cation layer of ≈ 0.28 nm; see Fig. 2 in the main text). After growth the films were moved in UHV to the analysis chamber for the STM and XPS measurements.

STM images were acquired at room temperature, in constant-current mode, and measuring empty states (positive sample bias $V_\text{sample}$). Electrochemically etched W tips were prepared by Ar$^+$ sputtering. Sometimes it helped to indent into the film or to apply current/voltage pulses. After such tip preparation steps, STM images were acquired after moving to a different spot on the sample.

XPS data were acquired in normal emission with a non-monochromatic Al K$\alpha$ source and a SPECS Phoibos 100 analyser. The areas of Mn 2$p$, La 4$d$, Sr 3$d$ XPS core-level peaks were evaluated with CasaXPS after subtracting a Shirley-type background.

Sputtering was performed with Ar$^+$ ions with 1 keV energy, 45° incidence, and sputter current density of ≈ 3 × 10$^{-5}$ A/cm$^2$, corresponding to ≈ 2 ions/nm$^2$ s (without correction for secondary electrons).

XRD data were collected at the TU Wien X-ray Centre [PANalytical Empyrean; Cu K$\alpha_1$ radiation obtained with a 2-pass Ge(220) hybrid monochromator and a 1/32° anti-divergence slit; a GaliPIX3D area detector with 0.02 rad Soller slits was used to measure reciprocal-space maps]. They were analysed with the xrayutilities Python package.[4]

For RBS, a 1.7 MeV $^4$He$^+$ ion beam was directed under normal incidence onto the sample, while the detector (resolution of 17 keV) was mounted under a backscattering angle of 170°. The total charge deposited was 10–20 μC (values determined by fitting the bulk signal for each spectrum). Areal densities and layer compositions were determined with the simulation code SIMNRA.

### S2. RBS and XRD characterization of atomically flat films

Figures S1 and S2 summarize the bulk characterization analysis (RBS and XRD) performed on cluster-free LSMO(110) films. RBS (Fig. S1) quantified the film stoichiometry as (La$_{0.78 \pm 0.03}$Sr$_{0.22 \pm 0.03}$)$_{1.06 \pm 0.05}$MnO$_3$, close to the composition of the target La$_{0.79 \pm 0.02}$Sr$_{0.21 \pm 0.02}$)$_{0.96 \pm 0.08}$MnO$_3$ (errors represent one standard deviation). The target composition was measured by inductively coupled plasma mass spectrometry – ICP-MS). Note that the bump indicated by the arrow in Fig. S1 results from the overlap of the La and Sr signals of LSMO. The experimental data (dots) were fitted (red solid line) by the sum of the individual components of the SrTiO$_3$ substrate and LSMO film.

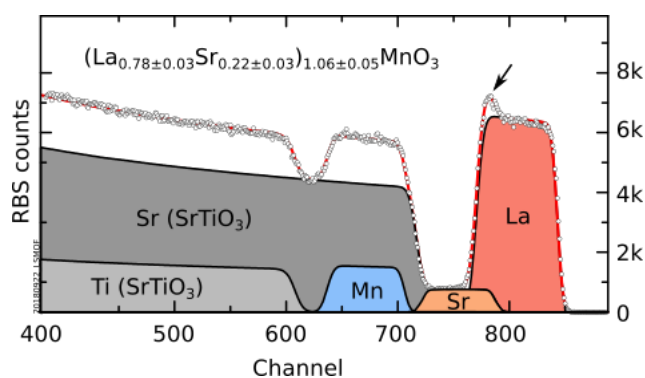

**Fig. S1.** RBS characterization of a La$_{0.8}$Sr$_{0.2}$MnO$_3$(110) film with 132 nm thickness grown with appropriate conditions (= 4.2 × 10$^{-2}$ mbar, and otherwise identical parameters as reported in the main text) to render an ideal surface morphology. Experimental data (dots) are fit well (red line) by the contribution of the SrTiO$_3$ substrate (grey) and LSMO (coloured). Fitting of the RBS spectrum yields a composition of (La$_{0.78 \pm 0.03}$Sr$_{0.22 \pm 0.03}$)$_{1.06 \pm 0.05}$MnO$_3$, close to the composition of the LSMO target, (La$_{0.79 \pm 0.02}$Sr$_{0.21 \pm 0.02}$)$_{0.96 \pm 0.08}$MnO$_{3.1}$. The bump indicated by the arrow does not correspond to accumulation of La at the interface; instead, it originates from the overlap of the La and Sr signals of LSMO.

It is worth point out that the RBS analysis is not sufficient to claim a perfect composition of the films, especially with the given error bars. This analysis supports our findings that the stoichiometry of the film fits in the expected range, and that no significant cation excesses are introduced into the bulk while the surface atomic structure remains stable at increasing thicknesses.

Note that the oxygen content cannot be reliably evaluated from RBS measurements. Other techniques (XPS, transmission electron microscopy – TEM, ICP-MS) were attempted but did not deliver reliable stoichiometry estimates. XPS suffered from forward-focusing effects and the lack of reliable reference samples for the single elements. In TEM, the preferential removal of Mn by $Ar^+$ milling while making the lamella produced different compositions as a function of the lamella's thickness. ICP-MS was problematic because of preferential cation dissolution and the presence of Sr in both film and substrate. Note that the oxygen stoichiometry of the films cannot be determined by RBS. Nonetheless, Kröger-Vink defect diagrams of 20%-doped LSMO[5] show that, at all experimental conditions presented in this work, the films should be oxygen-stoichiometric. At $10^{-6}$ mbar (the most reducing $O_2$ pressure used in our work) and 800 °C (i.e., 100 °C higher than our growth temperature; hence, more reducing), the concentration of oxygen vacancies per oxygen sites is at most $10^{-5}$.

Figure S2 shows XRD reciprocal-space maps measured for both symmetric (left column) and asymmetric reflexes. Asymmetric reflexes were measured with either the $[1\bar{1}0]$ or the $[00\bar{1}]$ directions of $SrTiO_3$ in the scattering plane, in both grazing incidence and grazing emission geometries. All panels show reflexes from the $SrTiO_3$ substrate (black dots) and the LSMO film (orange).

Multiple $SrTiO_3$ maxima are visible [see especially the $(22\bar{1})$ and $(221)$ reflexes], as already observed on other commercial single crystals.[3] LSMO reflexes are broader than those of $SrTiO_3$, as expected for heteroepitaxial films under slight stress[6] that relax by introducing misfit dislocations and forming mosaics.

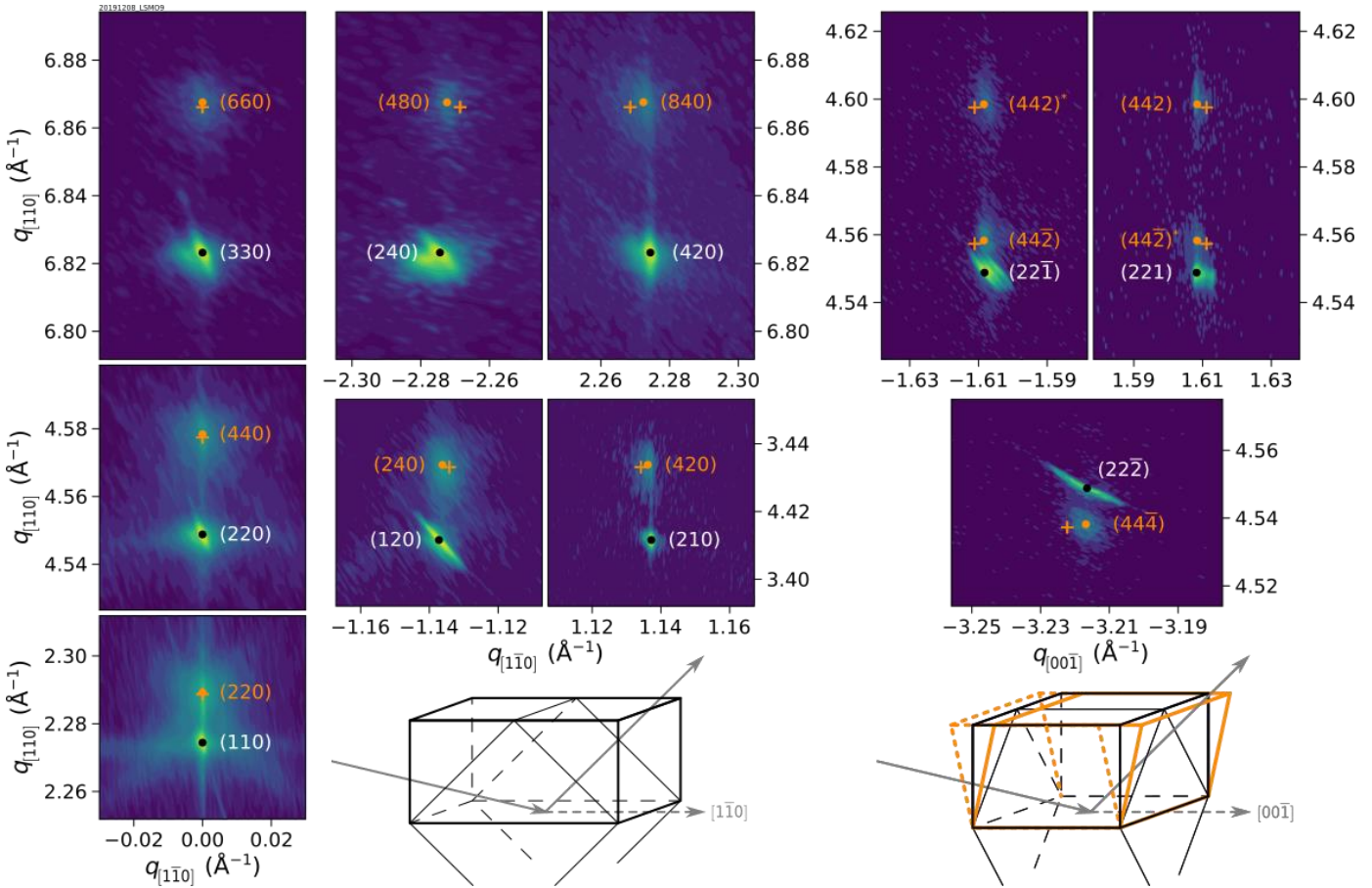

**Fig. S2.** XRD characterization of a 70 nm-thick $La_{0.8}Sr_{0.2}MnO_3$(110) film grown with appropriate conditions to obtain an ideal surface morphology (= $4.2 \times 10^{-2}$ mbar, and otherwise identical parameters as reported in the main text). Reciprocal-space maps acquired around symmetric (left column) and asymmetric (remaining panels) reflexes. Intensities are plotted with a logarithmic colour scale normalized to the intensity of the $SrTiO_3$ reflex in each panel. $q_{[110]}$ is the component of the momentum transfer perpendicular to the surface; $q_{[1\bar{1}0]}$ and $q_{[00\bar{1}]}$ are components in the surface plane. Negative in-plane components represent grazing-incidence configurations. $q = 2\pi/d$ when $d$ is the distance between diffracting planes. Black dots and white labels mark the position of reflexes of the $SrTiO_3$ substrate. Orange dots mark the position of reflexes of the best-fit structure for the LSMO film (see text). Orange crosses identify the positions for fully relaxed LSMO. Reflexes from two mirror-symmetric LSMO domains are visible in the rightmost panels (marked with an asterisk; also see the two domains schematically drawn in orange in the lower-right sketch). The measurement geometries are sketched at the bottom.

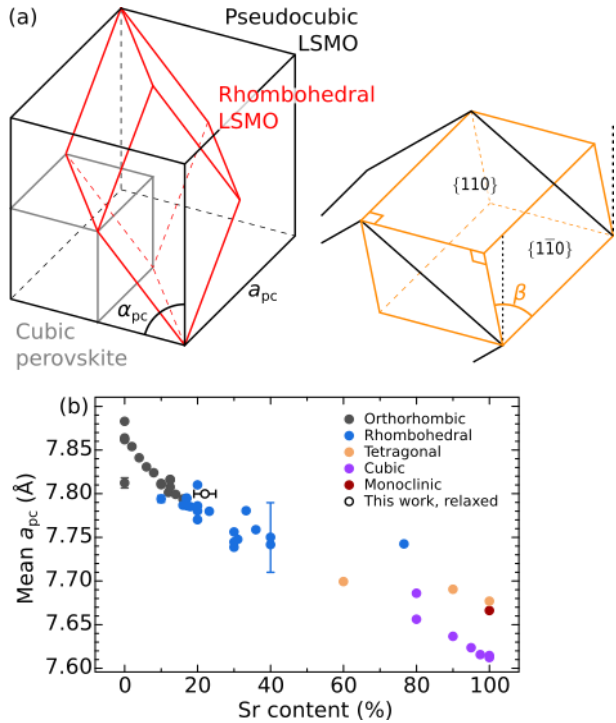

**Fig. S3.** (a) Sketches of the primitive (rhombohedral, red) and pseudocubic (black) unit cells of LSMO. In grey is the unit cell of a cubic perovskites such as SrTiO$_3$. In orange is the monoclinic unit cell bound by the $\{110\}$, and $\{1\bar{1}0\}$ planes of the pseudocubic unit cell. Notice that in this monoclinic cell, the $\{1\bar{1}0\}$ plane is non-rectangular. (b) Lattice constants of pseudocubic unit cells for LSMO as a function of the Sr content derived from the Crystallography Open Database.[11] For tetragonal, orthorhombic, and monoclinic primitive cells, the values plotted are a mean of the lattice parameters in the three directions.

Signs of this relaxation are also visible in the large-area STM images as long-range sub-unit-cell-height contrast modulations (see, e.g., Figs. 3b and 5b of the main text).

The raw XRD data were corrected for small offsets of the machine readings using the SrTiO$_3$ reflexes: the 2Θ offset was calculated by exploiting the known ratios of interplanar spacings for given $(hkl)$ pairs [e.g., the separation of (110) and (220) is one third of the separation of (110) and (330)];[7] the ω offset was derived from the known ratios of in- and out-of-plane components of reciprocal-space vectors $\mathbf{q}_{hkl}$. Both adjustments are independent of the lattice constants. To appropriately fit the offset-corrected SrTiO$_3$ reflexes, it was necessary to use a lattice constant of 3.90685(41) Å, deviating about 0.05% from the nominal constant of 3.905 Å.

To explain the LSMO reflexes, one must take the correct unit cell of LSMO. At the 80:20 La:Sr composition, LSMO has a rhombohedral primitive unit cell with $R\bar{3}c$ space group (red in Fig. S3a). When grown on a cubic perovskite like SrTiO$_3$, it is more convenient to transform this rhombohedral cell to a non-primitive pseudocubic (pc) one with side $a_{pc} \approx 8$Å and $\alpha_{pc} = 90° + \delta$ (black in Fig. S3a; approximately corresponding to a 2 × 2 × 2 supercell of the cubic perovskite unit cell, grey in Fig. S3a).[1,8] The $(hkl)$ indices for LSMO reported in Fig. S2 correspond to this pseudocubic unit cell. In the pseudocubic cell, the $\{110\}$ and $\{1\bar{1}0\}$ facets are inequivalent. The unit cell bound by the pseudocubic $\{110\}$ and $\{1\bar{1}0\}$ facets is monoclinic (orange in Fig. S3a) with $\cos\beta = \cos\alpha_{pc} / \cos(\alpha_{pc}/2)$. The $\{110\}$ plane has a rectangular 2D unit cell, while $\{1\bar{1}0\}$ is non-rectangular [angles $\beta$ and $(180° - \beta)$]. In principle, both orientations could be present after film growth. If the film grows $\{110\}$-oriented, the rectangular $\{110\}$ plane lies parallel to the surface of the SrTiO$_3$(110) substrate, and the out-of-plane lattice vector is oriented at an angle $\beta$. Conversely, $\{1\bar{1}0\}$-oriented films have their non-rectangular face on SrTiO$_3$(110), and the rectangular one perpendicular to the surface. The XRD data in Fig. S2 show that the film is purely $\{110\}$-oriented. This is evidenced by the panels in the right column of Fig. S2, showing a set of two reflexes from LSMO at the positions expected for $\{110\}$-oriented films (one marked with an asterisk). The presence of two reflexes, $(442)$ and $(44\bar{2})$, is expected because the out-of-plane tilt by $\beta$ breaks the mirror symmetry of SrTiO$_3$ and results in two symmetry-inequivalent domains. Instead, there would be only one set of spots for $\{1\bar{1}0\}$-oriented films roughly midway those in the right column of Fig. S2. The exclusive presence of $\{110\}$ — as opposed to the coexistence of $\{110\}$ and $\{1\bar{1}0\}$ — is reasonable since $\{1\bar{1}0\}$ should be energetically unfavourable: it requires additional shear stress to distort its in-plane non-rectangular cell and match the SrTiO$_3$(110) mesh.

To fit the LSMO reflexes, the following free independent parameters were considered: the lattice parameter $a_{pc}$ and angle $\alpha_{pc}$ of the unstrained (fully relaxed) pseudocubic cell, and the amount of relaxation of the misfit strain along the $[1\bar{1}0]$ and $[001]$ in-plane directions of SrTiO$_3$. For the fit, the relaxed pseudocubic LSMO cell was placed onto SrTiO$_3$(110); in-plane misfit strain was applied including relaxation; out-of-plane strain was derived using the known elastic constants of LSMO[9] following the approach of Ref. [10]. The distance in reciprocal space between experimental and calculated **q**-vectors was minimized using the differential evolution algorithm implemented in the scipy Python library.[11]

**Table S1.** Lattice constants (expressed in ångstroms) and monoclinic angles arising from the analysis of the XRD data.

|  | In-plane | | Out-of-plane | |
| --- | --- | --- | --- | --- |
|  | [001] | [1$\bar{1}$0] | [110] | $\beta(°)$ |
| Relaxed | 7.7998 | 5.5394 | 5.4910 | 90.714 |
| Strained | 7.8129 | 5.5301 | 5.4899 | 90.716 |
| SrTiO$_3$ | 7.8137 | 5.5251 | 5.5251 | 90.000 |

The fitting procedure yielded the results summarized in Table S1. The unstrained LSMO (orange crosses in Fig. S2) has a lattice constant of $a_{pc} = 7.7998$ Å and an angle of $\alpha_{pc} = 90.503°$. These values are in perfect agreement with the trend of lattice constants of LSMO as a function of the La:Sr ratio derived from the Crystallography Open Database (Fig. S3b).[12] The film is only partially relaxed (by 6% along $[001]$ and 34.7% along $[1\bar{1}0]$). This is also visible from the in-plane position of the reflexes in Fig. S2: Along $[001]$ (right column of Fig. S2), the LSMO periodicity closely matches the one of the SrTiO$_3$; along $[1\bar{1}0]$ (middle column), the LSMO reflexes are found at shorter in-plane reciprocal space positions, indicating an expansion of the LSMO lattice with respect to the substrate. The in-plane strain in the film relative to the relaxed structure has the same absolute value of ≈ 0.168% in the two directions. This corresponds to a very small deformation of the monoclinic cell, by less than 2 pm. It is tensile along $[001]$ and compressive along $[1\bar{1}0]$. In the out-of-plane $[110]$ direction, virtually no deformation is present (≈ −0.02%). It is likely that the largest relaxation occurs in correspondence of the boundary between mirror-symmetric domains. The direction of the topography modulations observed in large-area STM images fits with this interpretation.

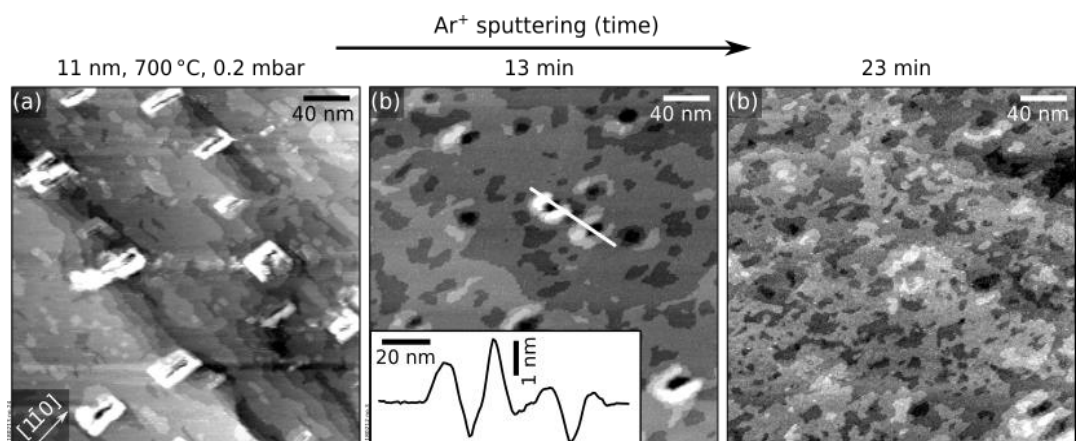

**Fig. S4.** Recovering ideal morphologies of an LSMO(110) film of 11 nm thickness by sputtering-annealing cycles (it is the same A-site rich film of Fig. 3c in the main text). (a–c) 300 × 300 nm² STM images, showing the flattening of the surface by Ar⁺ sputtering followed by O₂ annealing (> 45 min, 700 °C, 0.2 mbar). (b) After 13 min sputtering (≈1.5 × 10³ ions/nm²), the clusters have converted into pits with a depth of a few layers, surrounded by rims a few layers high (the inset shows the profiles of two pits). (c) After a total of 23 min (≈2.6 × 10³ ions/nm²) the surface is almost atomically flat (4 exposed layers).

## S3. UHV-based strategies to recover ideal surface morphologies

This section discusses UHV-based treatments to improve the morphology of rough films. First, consider surfaces such as in Figs. 3a, c of the main text, i.e., decorated by some clusters but otherwise showing large, atomically flat, terraces. Here, Ar⁺ sputtering plus high-pressure O₂ annealing can recover almost ideal surfaces. Figure S4 shows the morphology evolution of the film of Fig. S4a as a function of the sputtering time. (After each cycle, the sample was annealed for 1 h at 700 °C and 0.2 mbar O₂.) At the beginning, the surface is decorated by clusters (Fig. S4a). A closer inspection reveals that the clusters are made of square-edged rims that surround few-nanometre-deep holes. Sputtering for 13 min followed by annealing (Fig. S4b) removes most of the rims of the clusters and makes the pits more evident (see the line scan in the inset). The pits-plus-rims clusters are reminiscent of those formed during the Ti-rich homoepitaxy of SrTiO₃(110).[13] This supports previous findings[1] that nonstoichiometric growth causes pronounced sticking effects in both SrTiO₃(110) and LSMO(110). Further sputtering + annealing the surface (Fig. S4c) removes almost all the pits, yielding a flat surface with four layers exposed.

Now consider a surface like in Fig. S5a, formed by growing a film of 22 nm thickness at 1 × 10⁻² mbar O₂ and otherwise standard deposition parameters. Mn-rich clusters dominate the film morphology. Here, Ar⁺ sputtering plus high-pressure annealing does not recover a flat surface easily, see Figs. S5b–d. The number of exposed layers decreases with more cleaning cycles, but many pits remain (see Fig. S5c after 5 cycles). A longer anneal gives only a minor improvement (Fig. S5d). The most effective way to recover the surface is to anneal at more reducing conditions: Heating for 1 h at the same temperature but in UHV rather than in an O₂ background drastically reduces the number of exposed layers (Fig. S5e). A similar effect was reported for other oxide surfaces:[14] When the oxygen partial pressure (and, hence, its chemical potential) during annealing changes, reconstructions of different cation composition become more stable. The change of the surface reconstruction requires to move material across the surface to realize the thermodynamically stable composition. This helps to overcome kinetic barriers that otherwise prevent the formation of a smooth surface.

Note that sputtering always depletes the surface from Mn, shifting the surface towards the A-site rich end of LSMO(110) (see top row of Fig. 4 of the main text).[1] To recover the correct Mn stoichiometry, it is sufficient to deposit Mn plus high-pressure O₂ annealing.[1] After this step, the surface exhibits atomically flat terraces hundreds of nanometres wide (Fig. S5f).